\documentclass[aps,prd,twocolumn,floatfix,altaffilletter,superscriptaddress,preprintnumbers,
tightenlines,showpacs,showkeys,nofootinbib,notoccite]{revtex4-1}
\usepackage[utf8]{inputenc}
\usepackage[colorlinks=true,citecolor=blue,linkcolor=blue]{hyperref}
\usepackage[normalem]{ulem}
\usepackage{amsmath,amssymb, mathrsfs,esvect}
\usepackage{epsfig}
\usepackage{graphicx}               % Standard graphics package
\usepackage{url}
\usepackage{color}
\usepackage{slashed}
\usepackage{multirow}
\usepackage{placeins}
\usepackage[dvipsnames]{xcolor}
\usepackage{epstopdf}
\usepackage{soul}
\usepackage{tikz}
\usepackage[capitalise, english]{cleveref}
\usepackage{siunitx}
\usepackage{xspace}
\usepackage{booktabs}
\usepackage[capitalise]{cleveref}
\usetikzlibrary{trees}
\usetikzlibrary{decorations.pathmorphing}
\usetikzlibrary{decorations.markings}

\newcommand\myshade{80}
\colorlet{mylinkcolor}{ForestGreen}
\colorlet{mycitecolor}{Red}
\colorlet{myurlcolor}{violet}

\hypersetup{
	linkcolor  = mylinkcolor!\myshade!black,
	citecolor  = mycitecolor!\myshade!black,
	urlcolor   = myurlcolor!\myshade!black,
	colorlinks = true
}
\definecolor{jblue}{RGB}{20,50,100}
\definecolor{npurple}{RGB} {153, 51, 204}
\definecolor{wred}{RGB}{217,0,56}
\definecolor{white}{RGB}{255,255,255}
\definecolor{forestgreen}{HTML}{228B22}

\allowdisplaybreaks
\usepackage{tikz,xcolor,hyperref}

\definecolor{lime}{HTML}{A6CE39}
\DeclareRobustCommand{\orcidicon}{\hspace{-1mm}
	\begin{tikzpicture}
		\draw[lime, fill=lime] (0,0) 
		circle [radius=0.16] 
		node[white] {{\fontfamily{qag}\selectfont \tiny \,ID}};
		\draw[white, fill=white] (-0.0525,0.095) 
		circle [radius=0.007];
	\end{tikzpicture}
	\hspace{-3mm}
}

\foreach \x in {A, ..., Z}{\expandafter\xdef\csname orcid\x\endcsname{\noexpand\href{https://orcid.org/\csname orcidauthor\x\endcsname}
		{\noexpand\orcidicon}}
}

\newcommand{\prn}[1]{ \left(  #1 \right) }

\usepackage{tikz}
\usepackage{tikz-feynman}
\tikzfeynmanset{compat=1.0.0}

\begin{document}
%%%%%%%%%%%%%%%%%%%%%%%%%%%%%%%%%%%%%%%%%%%%%%%%%%
	
\title{Dark Matter Catalyzed Baryon Destruction}
	\author{Yohei Ema\orcidA}
	\email{ema00001@umn.edu}
	\affiliation{William I. Fine Theoretical Physics Institute, School of Physics and Astronomy\\ University of Minnesota, Minneapolis, MN 55455, USA}
	\affiliation{School of Physics and Astronomy, University of Minnesota, Minneapolis, MN 55455, USA}
	\author{Robert McGehee\orcidB}
	\email{rmcgehee@umn.edu}
	\affiliation{William I. Fine Theoretical Physics Institute, School of Physics and Astronomy\\ University of Minnesota, Minneapolis, MN 55455, USA}
	\author{Maxim Pospelov}
	\email{pospelov@umn.edu}
		\affiliation{William I. Fine Theoretical Physics Institute, School of Physics and Astronomy\\ University of Minnesota, Minneapolis, MN 55455, USA}
	\affiliation{School of Physics and Astronomy, University of Minnesota, Minneapolis, MN 55455, USA}
	\author{Anupam Ray\orcidD}
	\email{anupam.ray@berkeley.edu}
	\affiliation{School of Physics and Astronomy, University of Minnesota, Minneapolis, MN 55455, USA}
 \affiliation{Department of Physics, University of California Berkeley, Berkeley, California 94720, USA}
\date{\today}
	
%%%%%%%%%%%%%%%%%%%%%%%%%%%%%%%%%%%%%%%%%%%%%%%%%%%%%%%%%%%%%%%%%%%
\begin{abstract}
WIMP-type dark matter may have additional interactions that break baryon number, leading to induced nucleon decays which are subject to direct experimental constraints from proton decay experiments. In this work, we analyze the possibility of continuous baryon destruction, deriving strong limits from the dark matter accumulating inside old neutron stars, as such a process leads to excess heat generation. We construct the simplest particle dark matter model that breaks the baryon and lepton numbers separately but conserves $B-L$. Virtual exchange by DM particles in this model results in di-nucleon decay via $nn\to n\bar\nu$ and $np\to ne^+$ processes.
\end{abstract}
\maketitle
\preprint{FTPI-MINN-24-12, UMN-TH-4320/24, N3AS-24-024}

%%%%%%%%%%%%%%%%%%%%%%%%%%%%%%%%%%%%%%%%%%%%%%%%%%
\section{Introduction}
%%%%%%%%%%%%%%%%%%%%%%%%%%%%%%%%%%%%%%%%%%%%%%%%%%
The Standard Model (SM) of particle physics has done a remarkable job of understanding the subatomic cosmos as it exists today. Despite its immense success, there are a number of key concerns, and one such concern is the observed excess of matter to antimatter which essentially requires baryon-number-violating (BNV) interactions~\cite{Sakharov:1967dj}.
Baryogenesis is the dynamical process of generating the preponderance of matter over antimatter, and several scenarios for it exist within various extensions of the SM. On the experimental front, no observations of the BNV interactions in the laboratory have been achieved, resulting in, {\em e.g.},  tight upper bounds on the  lifetime of the proton and the neutron-antineutron oscillation. Remarkably, the SM thermal processes during the electroweak epoch of the early Universe are believed to greatly enhance the BNV processes without contradicting stringent limits from proton decay~\cite{Kuzmin:1985mm,Koren:2022bam}. 

Dark matter (DM) represents another great mystery, as its identity remains unknown, despite a variety of experimental and theoretical efforts. The closeness of the DM and baryon energy densities (within a factor of $\sim 5$) has generated some speculation that  DM-genesis and baryogenesis may in fact be related (see {\em e.g.} Refs.~\cite{Kaplan:2009ag,Shelton:2010ta,Davoudiasl:2010am,Cui:2011qe,Bringmann:2018sbs,Elor:2020tkc} for a representative set of ideas). Popular scenarios include assigning some dark particles a baryon number, so that the process of baryogenesis may be thought of as the process of secluding the baryon number inside the SM sector and the anti-baryon number in the DM sector \cite{Davoudiasl:2010am}. 

In this paper, we would like to take a step back from concrete scenarios of baryogenesis, and ask a question: What if DM had BNV interactions?\footnote{For other ideas connecting nucleon-number-changing processes with dark sector physics, see {\em e.g.} Refs.~\cite{Fornal:2018eol,Goldman:2019dbq,Johns:2020mmo,Johns:2020rtp,Berezhiani:2020zck,McKeen:2021jbh,Koren:2022axd,Davoudiasl:2023peu}.}
This opens an interesting possibility of $\mathcal{O}(\rm GeV)$ energy release in interactions of DM particles with baryons. This is in contrast with the elastic scattering of baryons and DM particles, where the energy release is typically limited to tens of keV or less. The goal of our paper is to explore a continuous process of DM-catalyzed destruction of the baryon number. In the past, several studies addressed the phenomenological consequences \cite{Kuzmin:1983by,Arafune:1983sk,Kolb:1984yw} of grand unified theory (GUT) monopole-induced breaking of the baryon number \cite{Callan:1982au,Rubakov:1982fp}. While GUT monopole masses are limited to scales of $\mathcal{O}(10^{16} \text{ GeV})$, a generic DM model can have an arbitrary mass scale. In this paper, we focus on DM composed of weakly interacting massive particles (WIMPs), and we endow them with small BNV interactions. Since the WIMP abundance can be many orders of magnitude larger than that of GUT monopoles, BNV signatures can be far more pronounced.

For example, the WIMP BNV interactions can play a significant role in the early Universe at temperatures above the WIMP mass when its abundance is thermal. At these temperatures, other BNV interactions are active: electroweak sphalerons break $B+L$ while conserving $B-L$. It is well understood that if other processes break $B-L$, any pre-existing baryon asymmetry may be completely washed out~\cite{Campbell:1990fa,Campbell:1992jd}. 

While the threat of total baryon number erasure is known to have caveats~\cite{Dreiner:1992vm}, we use it to limit the scope of our paper to $(B-L)$-preserving BNV interactions. In this case, the symmetry breaking pattern of DM-baryon interactions is the same as that of the sphalerons, and many conventional baryogenesis scenarios based on non-zero $B-L$ asymmetry (such as leptogenesis~\cite{Fukugita:1986hr}) will work without major complications. Therefore, we will not consider $\bar{n}-n$ oscillations, $nn$ annihilation to pions, or other BNV processes that violate $B-L$.

\begin{figure}[t]
\includegraphics[width=0.4\textwidth]{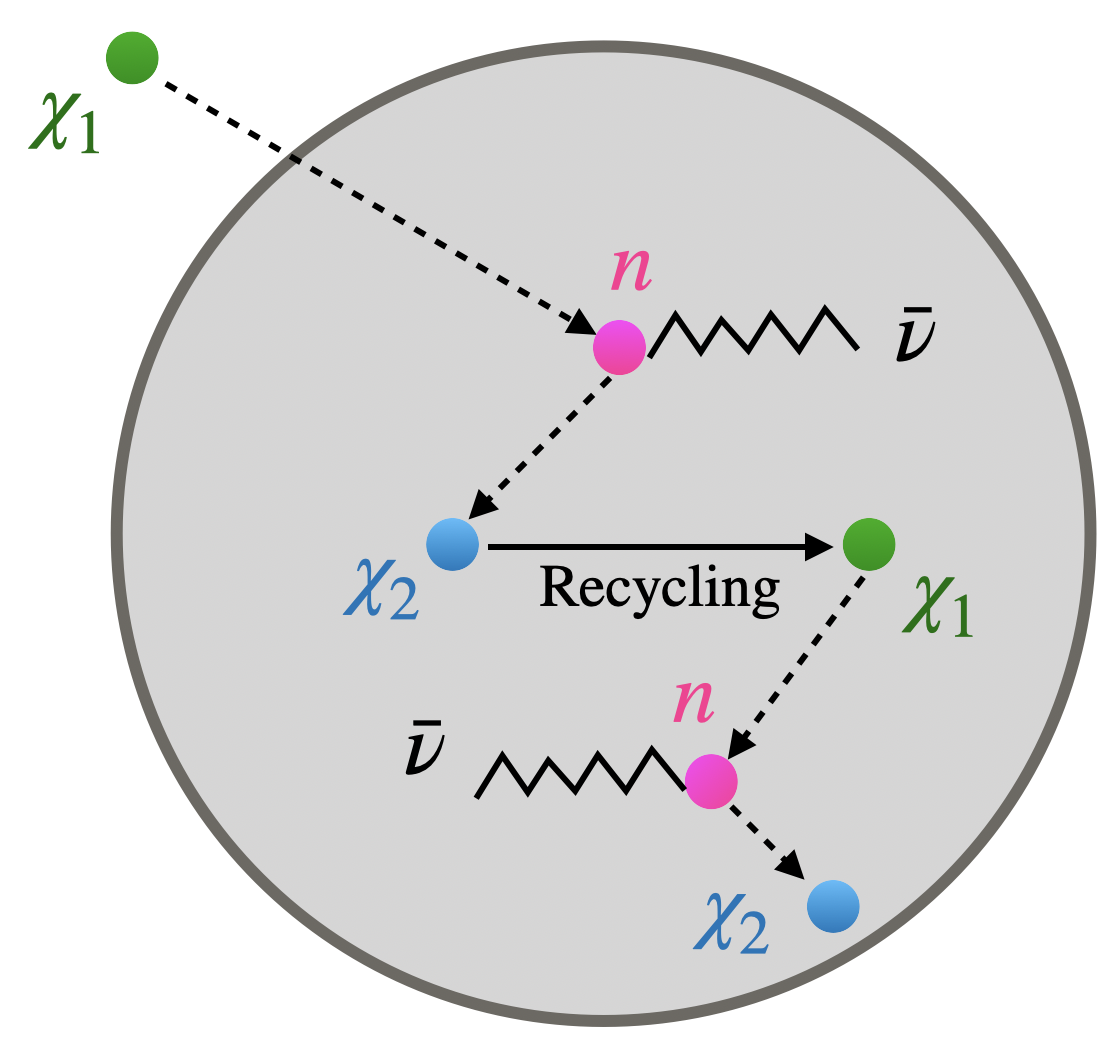}
\caption{Schematic diagram for DM-catalyzed baryon destruction inside a NS.}
\label{fig:DMinNS}
\end{figure}

One of the key consequences of such BNV processes is nucleon destruction: $\chi+N \to \chi+$ Energy.  More specifically, if a DM particle $\chi$ could destroy nucleons (while preserving $B-L$),
\begin{equation}
\label{eq:DMinducedBNV}
\chi + n \to \chi + \bar{\nu} \,\,~\textrm{and/or}~\,\, \chi + p \to \chi + e^+,
\end{equation}
then the signals we might see today would be striking. Incoming DM could trigger $\mathcal{O}(\text{GeV})$ energy releases inside large-volume neutrino detectors such as Super-Kamiokande (SK)~\cite{Super-Kamiokande:2020bov}. Captured DM particles could also catalyze a continuous energy release in old, cool neutron stars (NSs)
as schematically shown in Fig.~\ref{fig:DMinNS}. 
Moreover, the presence of BNV interactions may have consequences for processes without on-shell DM particles.
Even if kinematics forbids tree-level proton decays, the presence of BNV interactions could allow higher-loop proton decays through virtual DM particles. 

For concreteness, we consider two DM species $\chi_1$ and $\chi_2$ with $\chi_2$ slightly heavier but nearly degenerate masses (to avoid stringent terrestrial BNV constraints; see Sec.~\ref{subsec:di-nucleon}). Because of the BNV interactions, $\chi_1$ can destroy nucleons via $\chi_1+p/n \to \chi_2+e^+/\bar{\nu}$, and a) result in a novel heating mechanism in cold NSs and b) release a detectable amount of energy inside large-volume neutrino detectors, such as SK. In order to continue this process, $\chi_2$ needs to decay/oscillate back to $\chi_1$ on a relatively short timescale. Oscillations, for example, can be realized via $\chi_1-\chi_2$ mass mixing. In this way, $\chi_{2}$ gets effectively ``recycled", {\em i.e.} efficiently converts back to $\chi_{1}$, and the nucleon destruction via BNV processes continues.\footnote{For this recycling to occur, it is crucial to have an actual BNV interaction. This is in contrast to the model in, \emph{e.g.}, Ref.~\cite{Huang:2013xfa}, where one can assign a baryon charge to the dark sector so that the total baryon number is conserved.} 
We emphasize that this oscillation is not the only option; we can instead have
a decay of $\chi_2$ back to $\chi_1$. 
To make our general discussion independent of specific choices,
we work in the oscillation basis, not in the mass basis, in Sec.~\ref{sec:estimate}.

In this paper, we explore each of these striking signals of BNV DM in turn. 
Sec.~\ref{sec:estimate} describes the main physics idea, without specifying the model details.
In Sec.~\ref{subsec:SK}, we estimate the approximate bound coming from neutron destruction in SK.
In Sec.~\ref{subsec:NS}, we calculate the constraints coming from heating NSs due to the DM-catalyzed neutron destruction. 
There we also summarize the basics of DM capture in NSs.
 
Next, we detail a simple toy model of DM in Section~\ref{sec:model} as a concrete realization of our idea.
The toy model allows us a concrete comparison of the constraints derived in Sec.~\ref{sec:estimate}
with terrestrial constraints on BNV processes where DM particles appear virtually in the loop.
In particular, we see that the introduction of two components, $\chi_1$ and $\chi_2$, allows us to avoid stringent terrestrial BNV constraints while preserving interesting signals inside NSs.

Finally, we conclude with some discussions of other possible unusual DM interactions in Sec.~\ref{sec:disc}.

%%%%%%%%%%%%%%%%%%%%%%%%%%%%%%%%%%%%%%%%%%%%%%%%%%
\section{Estimate of Super-Kamiokande Constraint and Neutron Star Heating}
\label{sec:estimate}
%%%%%%%%%%%%%%%%%%%%%%%%%%%%%%%%%%%%%%%%%%%%%%%%%%
\subsection{Constraint from Super-Kamiokande}
\label{subsec:SK}

The promising signals of DM-induced BNV interactions inside large-volume neutrino detectors such as SK have long been recognized~\cite{Davoudiasl:2011fj,Demidov:2015bea}. While these same interactions could occur in large dark matter experiments, they are always above threshold in the much-larger neutrino detectors, which are thus more constraining. In our model, the following processes can occur inside the SK fiducial volume:
\begin{eqnarray}
\label{eq:BNV_pi0}
	\chi_1 +p \to \chi_2 +e^+,~~~~~~~\\
	%\,\,~\rm{and}~\,\, \chi_2 \to \chi_1\\
 \chi_1 +p \to \chi_2 +e^+ + (1\mbox{-\,to\,-}6)\pi.
 \label{eq:withpions}
\end{eqnarray}
Any process that annihilates a proton may also lead to the emission of pions.\footnote{The process in Eq.~\eqref{eq:withpions} is limited to six pions simply due to kinematics.} We concentrate on the pion-less process for simplicity, noting that the cross sections for Eq.~\eqref{eq:withpions} could be comparable to Eq.~\eqref{eq:BNV_pi0}. 
Without yet specifying any DM model, we denote the cross section times the relative velocity for such process as $v\sigma_\text{BNV}$. Then the rate of these positron-producing events in SK is
\begin{equation}\label{skrate}
		R_{\rm SK} = \frac{\rho_\chi}{m_\chi}\times v\sigma_\text{BNV}\times N_p^\text{SK},
\end{equation}
where $\rho_\chi = 0.4 \text{ GeV}/\text{cm}^3$ is the local DM energy density, $N_p^\text{SK} = \left(5/9\right) \times \left(22.5\,\textrm{kT}/m_n\right)$ is the total number of protons inside the fiducial volume of SK~\cite{Super-Kamiokande:2020bov}, and $m_n$ is the mass of the nucleon. If neutrons are considered as initial targets, either the antineutrino is emitted as in Eq.~\eqref{eq:DMinducedBNV} or the charge is compensated by the additional pion release, $\chi_1 +n \to \chi_2 +e^++\pi^-$. The former is less interesting in the context of a SK signal, while the latter has a somewhat reduced phase space. However, it has a more symmetric energy deposition, leading to less background, and will be constrained by the most sensitive nucleon-decay searches. 

The proton's conversion to a positron [Eq.~\eqref{eq:BNV_pi0}] is very similar to the electron-like event due to the charged current interaction of atmospheric neutrinos $(\nu_e+\bar{\nu}_e)$. The latter is a well-established signal, with a rate of $\sim$\,2 events/day~\cite{Super-Kamiokande:2015qek}, and less than $\mathcal{O}(1)$ event per day if a proper energy window around $m_p$ is selected. For the most conservative estimate, one may simply compare the rate in Eq.~\eqref{skrate} to 2 events/day.
Of course, an additional $\pi^0$ $(\chi_1 +p \to \chi_2 + \pi^0+e^+)$ would result in more symmetric events with three electron-like rings reconstructing $m_p$ and significantly fewer background events. Assuming the background is $\sim 25 \times$ lower for events with the $\pi^0$, we can set a very tight limit by demanding fewer than 30 events/year (as shown in Fig.~\ref{fig:results}).\footnote{This is aggressive since by requiring an additional $\pi^0$ in the final state, we expect that the cross section gets suppressed compared to the original $\sigma_\mathrm{BNV}$ of the process~\eqref{eq:BNV_pi0}; see, {\em e.g.}, the analysis of Ref. \cite{Huang:2013xfa}.} However, we see below that, even with this rather aggressive treatment of the SK bound, the NS heating (discussed in the next sub-section) provides a much greater sensitivity to the BNV interactions. Therefore, we primarily focus on probing BNV interactions via NS heating. 
\begin{figure}
\includegraphics[width=0.47\textwidth]{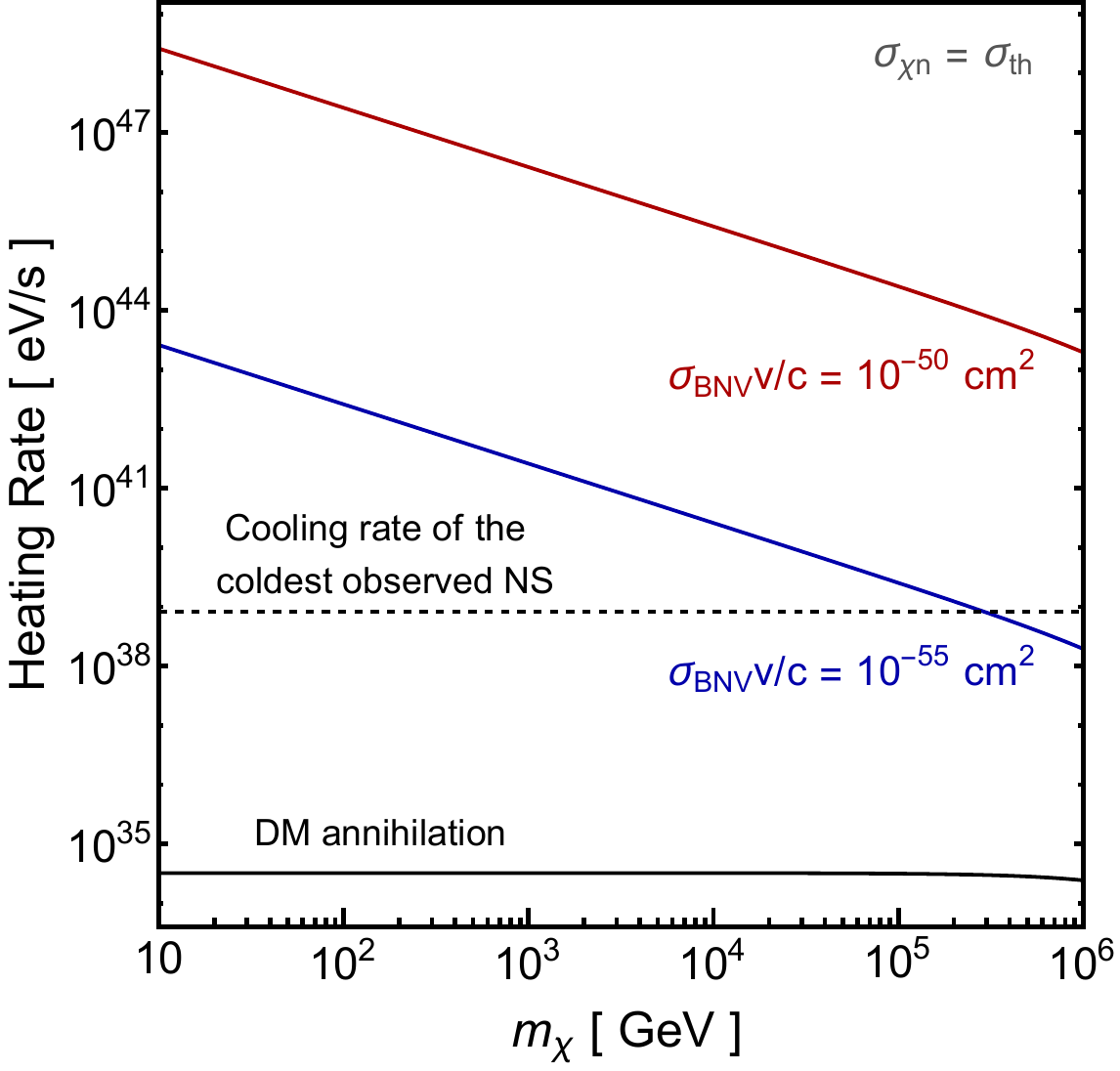}
\caption{The heating rates from the annihilation of accumulated DM inside a NS and DM-catalyzed baryon destruction with $\sigma v_{\rm BNV}/c = 10^{-50}$\,cm$^2$ and $10^{-55}$\,cm$^2$. Since the total amount of DM captured inside a NS is minuscule compared to the total mass of the neutrons, the heating rate from DM annihilation is significantly smaller. We use the typical NS parameters (see text for details) to estimate the heating rates and the DM-nucleon scattering cross section is taken as $\sigma_{\rm th} = 2.3 \times 10^{-45} \,\mathrm{cm}^2$ to achieve the maximal accumulation. We note that $M_*/t_* = 1.7\times 10^{50}\,\mathrm{eV}/\mathrm{s}$ for PSR J2144-3933 so that a larger $\sigma_\mathrm{BNV}v/c$ would even cause an $\mathcal{O}(1)$ destruction of the NS within its lifetime.}
\label{fig:heating}
\end{figure}
%%%%%%%%%%%%%%%%%%%%%%%%%%%%%%%%%%%%%%%%%%%%%%%%%%
\subsection{Constraint from Neutron Star Heating}
\label{subsec:NS}
%%%%%%%%%%%%%%%%%%%%%%%%%%%%%%%%%%%%%%%%%%%%%%%%%%
The flux of galactic halo DM particles through astrophysical bodies such as stars can result in DM accumulation due to occasional interactions with the stellar material~\cite{1985ApJ...296..679P,Gould:1987ir,Goldman:1989nd,Bertone:2007ae}.
In the optically thin regime (for small DM-nucleon scattering cross sections), compact stars such as NSs are the ideal targets for DM searches as they can capture significant amounts of DM particles despite small cross sections. This is simply because, in this regime, capture primarily occurs via single collisions, and the largest column density of nucleons can be encountered in a NS, resulting in a very efficient capture process, even for a relatively small scattering cross section. Quantitatively, a solar-mass NS (residing in the Solar neighborhood) with a typical radius of 10 km can capture DM particles $\sim 7 \times 10^4$  times more efficiently than the Sun for the same DM-nucleon interactions.

If DM has BNV interactions, the captured DM particles inside a NS can destroy the neutrons to yield neutrinos, liberating heat, and resulting in a novel heating mechanism of cold NSs. More specifically, we are interested in the following process:
\begin{equation}
	\chi_1 +n \to \chi_2 + \bar{\nu}\,\,\rm{and}\,\, \chi_2 \to \chi_1.
\end{equation}
We assume that the mass difference between $\chi_1$ and $\chi_2$ is negligible compared to $m_n$ so that the neutrino has energy $\sim m_n$.
Since NSs are opaque to neutrinos of GeV-scale energy, the liberated energy is consumed by the NSs and heats them up.
Since we expect that the main consequence of BNV interactions is catalyzed heat production in NSs, the coldest NS observed to date, PSR J2144-3933~\cite{Guillot:2019ugf}, potentially has the most constraining power. For this particular NS, we assume its radius ($R_{\star})= 11$ km, mass ($M_{\star}) = 1.4 M_{\odot}$, surface temperature ($T_{\star}) = 2.85$ eV ($T_{\star}$ is reported to be $\leq$ 2.85 eV ; we take the largest value to be conservative), and lifetime ($t_{\star}$) $\approx$ 300 Myr~\cite{Guillot:2019ugf,Raj:2024kjq}.

To estimate the heat generation, we first calculate the number of DM particles captured over the lifetime of the NS. GeV-PeV DM particles get trapped inside the NS after a single collision~\cite{Bramante:2017xlb,Bhattacharya:2023stq}. Therefore, if the DM particles do not annihilate among themselves, the number of DM particles captured in the NS  is~\cite{Bramante:2017xlb}
\begin{equation}
	N_{\chi_1} = \epsilon_{\rm cap} \sqrt{\frac{6}{\pi}} \frac{\rho_{\chi}}{m_{\chi}} \pi R_{\star}^2\,\bar{v}\, \frac{v^2_{\rm esc}}{\bar{v}^2}\,\left(1-\frac{1-e^{-A^2}}{A^2}\right)  t_{\star}\,,
	\end{equation}
where $\epsilon_{\rm cap} $ = Min\,$ \left[1,{\sigma_{\chi n}}/{\sigma_{\rm th}} \right]$ is the capture efficiency which depends on the DM-nucleon scattering cross sections ($\sigma_{\chi n}$). $\sigma_{\rm th} = \pi R^2_{\star}/N_n  = 2.3 \times 10^{-45}$ cm$^2$ denotes the threshold cross section up to which the single-collision approximation is valid. For $m_{\chi} \leq 10^6$ GeV, the threshold cross section also implies that all of the transiting DM particles get trapped, and the geometric capture limit is reached. We assume that the average velocity of the DM particles in the Galactic halo is $(\bar{v})$ = 220 km/s, and the ambient DM density in the vicinity of the NS is $(\rho_{\chi}) =0.4$ GeV/cm$^3$. The escape velocity at the NS surface is taken as $v_{\rm esc}=1.8 \times 10^5$ km/s, and the dimensionless factor involving $A^2=6m_{\chi}m_n v^2_{\rm esc}/\bar{v}^2 (m_{\chi} - m_n)^2$ accounts for inefficient momentum transfers in the DM-nucleon scattering. For DM masses below $\sim 10^6$ GeV, the dimensionless factor involving $A^2$ evaluates to unity, and as a result, the number of captured DM particles scales inversely with the DM mass. However, for $m_{\chi} \geq  10^6$ GeV, the kinematic suppression becomes important, and the factor evaluates to $A^2/2$, leading to $N_{\chi_1} \propto 1/m^2_{\chi}$. We neglect possible general-relativistic corrections of the capture rate, which can enhance the capture rate by at most a factor of 2~\cite{Kouvaris:2007ay}.

The heating rate from neutron destruction is
\begin{equation}
\label{eq:heatrate}
	\frac{dE_{\rm heat}}{dt} = N_{\chi_1} \times v \sigma_\mathrm{BNV}\times n_n m_n\,,
\end{equation}
where $n_n$ is the neutron number density inside the NS. While a more careful analysis would account for the non-uniform density of neutrons within NSs, we assume that a uniform neutron density in Eq.~\eqref{eq:heatrate} is sufficiently accurate to capture the physics for this initial study. Notice that this novel heating mechanism inside NSs is drastically different from the heating via captured DM annihilation~\cite{Kouvaris:2007ay,Bertone:2007ae,Dasgupta:2020dik} or kinetic energy transfer~\cite{Baryakhtar:2017dbj,Raj:2017wrv}. In those cases, the energy injection is limited by the total energy density of the DM accumulated inside the NS, which is significantly smaller. For $m_\chi = 100 \text{ GeV}$ and $\sigma_{\chi n} = \sigma_{\rm th}$,
the heating rate via DM annihilation (kinetic energy transfer) is $\sim 3.2 \times 10^{34}$ ($\sim 0.9 \times 10^{34}$) eV/s. For DM-catalyzed nucleon destruction, the heating rate scales linearly with the BNV interaction strength and is $\sim 2.6 \times 10^{47}$ eV/s for $\sigma_{\rm BNV} v/c=10^{-50}$ cm$^2$; see Fig.~\ref{fig:heating}. This relatively large heating rate simply arises from the fact that in the BNV scenario the energy is provided by the mass of the neutrons, and neutrons are much more abundant than the captured DM particles. Typically, a NS can accumulate a maximum of $\mathcal{O} (10^{-16})\,M_{\odot}$ mass throughout its lifetime which is significantly smaller than the total mass of the neutrons. 

\begin{figure}
\includegraphics[width=0.47\textwidth]{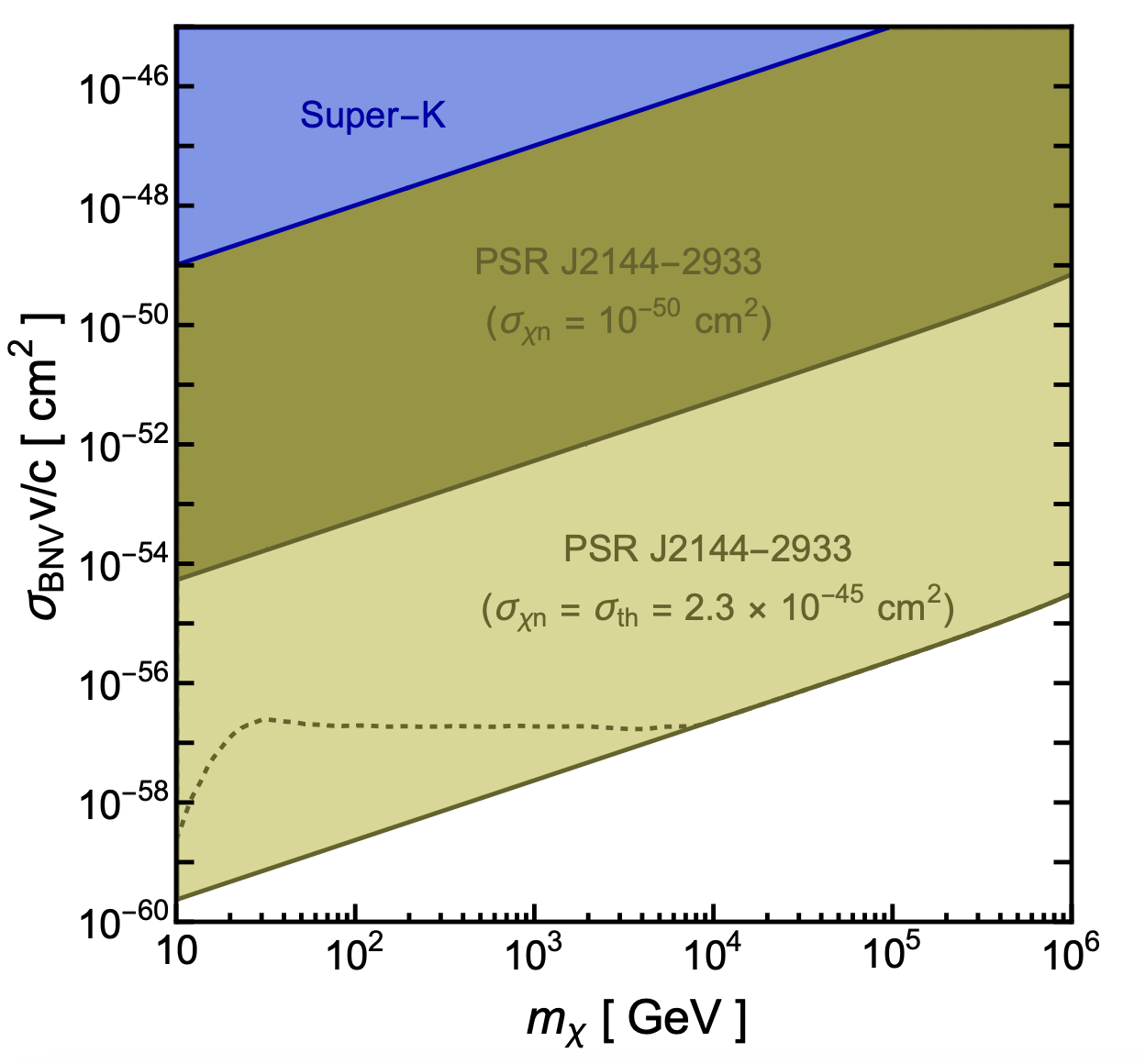}
\caption{Constraints on DM BNV interactions from the non-observation of anomalous heating of the cold NS PSR J2144-2933 (yellow shaded regions) for different DM-nucleon scattering cross sections. Since the accumulation rate scales linearly with the DM-nucleon scattering cross sections, constraints on BNV interactions become stronger with larger $\sigma_{\chi n}$, and maximal sensitivity can be achieved for $\sigma_{\chi n} = \sigma_{\rm th}$ (the geometric accumulation rate). The dashed yellow curve corresponds to $\sigma_{\chi n}$ saturating the current direct-detection bound~\cite{LZ:2022lsv}. Constraints from Super-Kamiokande (see text for more details) are also shown by the blue shaded region.}
\label{fig:results}
\end{figure}

Old NSs can be treated as black bodies that cool according to the classical Stefan-Boltzmann law~\cite{Yakovlev:2004iq}. Therefore, their cooling rate may be approximated by
\begin{align}	
	\frac{dE_{\rm loss}}{dt} &= 4\pi R_\star^2 \sigma_{\rm SB} T_\star^4 
	\nonumber \\
	&= 6.4 \times 10^{38} \frac{\text{eV}}{\text{s}} \prn{\frac{R_\star}{11\,\mathrm{km}}}^2 \prn{\frac{T_\star}{2.85\,\mathrm{eV}}}^4,
\end{align}
where $\sigma_{\rm SB}$ is the Stefan-Boltzmann constant. Here, we stress that, even with the coldest observed NS, the heating rate via DM annihilation (or via kinetic energy transfer) is not sufficient to induce any observable effect, whereas, in the case of DM-catalyzed baryon destruction, the heating rate causes observable temperature increases in cold NSs (see Fig.~2). Therefore, the non-observation of any anomalous heating of cold NSs provides a novel way of probing DM BNV interactions. We obtain the constraint on BNV interactions shown in Fig.~\ref{fig:results} by simply demanding that $dE_{\rm heat}/dt \leq dE_{\rm loss}/dt$.
%%%%%%%%%%%%%%%%%%%%%%%%%%%%%%%%%%%%%%%%%%%%%%%%%%
\section{A Simple Toy Model}
\label{sec:model}
%%%%%%%%%%%%%%%%%%%%%%%%%%%%%%%%%%%%%%%%%%%%%%%%%%
While one could frame the discussion in terms of the overall DM-nucleon BNV cross sections, in this section, we would like to take this topic further, and construct a simple model that realizes DM BNV physics. The model will give us some idea of whether the experimental/observational sensitivity derived in the previous section is actually realistic in terms of model parameters. Moreover, a concrete model allows comparison between on-shell DM scattering with BNV processes where DM particles appear virtually in the loop processes. 

We now describe a concrete realization of the above physics. We would like to emphasize that the model below is not unique, but is rather one in a wide family of possibilities, chosen here for its relative simplicity.   
There are three key ingredients:
(1) BNV interactions between nucleons and DM,
(2) elastic interactions between nucleons and DM to capture DM efficiently in NSs,\footnote{We assume that the DM is asymmetric in the current universe
to avoid annihilations induced by $\mathcal{L}_2$ in NSs.}
and (3) a mass mixing between $\chi_1$ and $\chi_2$ to recycle $\chi_1$'s in NSs.
These ingredients are provided by corresponding terms in our toy model's Lagrangian:
\begin{align}
	\mathcal{L} = \mathcal{L}_1 + \mathcal{L}_2 + \mathcal{L}_3,
\label{eq:Lag}
\end{align}
where at the effective field theory level we have
\begin{align}
\label{eq:LBNV}
	\mathcal{L}_1 &= G_{\mathrm{BNV}} \bar{\chi}_2\gamma_\mu \chi_1\times
	\left(\overline{e^+} \gamma^\mu p + \overline{\bar{\nu}}\gamma^\mu n \right)
	+ (\mathrm{h.c.}),
	\\
	\mathcal{L}_2 &= G_\chi \left(\bar{\chi}_1 \gamma_\mu \chi_1 + \bar{\chi}_2 \gamma_\mu \chi_2\right)
	\left(\bar{p} \gamma^\mu p + \bar{n}\gamma^\mu n\right),
	\\
	\mathcal{L}_3 &= - \frac{\Delta m_\chi}{2} \bar{\chi}_2 \chi_1 + (\mathrm{h.c.}).
\end{align}
For simplicity, we assume isospin singlet couplings between DM and nucleons
and a universal coupling of $\chi_1$ and $\chi_2$ to the nucleons in $\mathcal{L}_2$.
The interactions in $\mathcal{L}_1$ by themselves do not break baryon number
(or, more precisely, $B+L$)
as we can assign baryon charges to $\chi_1$ and $\chi_2$ that differ by unity. This is broken by the mass mixing term in $\mathcal{L}_3$, indicating that any BNV processes that do not involve $\chi_i$'s in the initial or final states are suppressed 
by the small mass mixing $\Delta m_\chi/m_\chi$.
Indeed, the main motivation for introducing two components, $\chi_1$ and $\chi_2$,
is to avoid stringent constraints from di-nucleon decay to a nucleon plus a lepton, as we see below.
In the following, we take the masses of $\chi_1$ and $\chi_2$ equal, $m_\chi$, except
for the small mass mixing arising from $\mathcal{L}_3$.
We focus on the case $m_\chi \gtrsim m_n$ to prohibit the nucleon decay $N \to \bar{\chi}_1 + \chi_2 + e^+/\bar{\nu}$.

We stress again that oscillations are not the only possibility to convert $\chi_2$ back to $\chi_1$.
A light scalar field $S$ could permit the decay $\chi_2 \to \chi_1 + S$. The neutron could then also decay at one-loop via $n \to S + \bar{\nu}$. 
We have verified that these neutron decays could be sufficiently slow and the $\chi_2$ decays sufficiently fast, but we will not pursue further discussion of models based on the $\chi_2\to \chi_1$ decay. 

%%%%%%%%%%%%%%%%%%%%%%%%%%%%%%%%%%%%%%%%%%%%%%%%%%
\subsection{Reaction rates}
%%%%%%%%%%%%%%%%%%%%%%%%%%%%%%%%%%%%%%%%%%%%%%%%%%
We now list the reaction rates expressed in terms of the model parameters. 
The BNV cross section induced by DM, $\chi_1 n \to \chi_2 \bar{\nu}$, is given by
\begin{align}
	\sigma_\mathrm{BNV} v/c 
	&= \frac{G_\mathrm{BNV}^2 m_n^2}{32\pi}
	\frac{(2m_\chi + m_n)^2(2m_\chi^2 + 4 m_\chi m_n + 3m_n^2)}{(m_\chi + m_n)^4},
\end{align}
in the non-relativistic limit of the initial particles.
In the limit $m_\chi \gg m_n$, it reduces to
\begin{align}
	\sigma_\mathrm{BNV} v/c = \frac{G_\mathrm{BNV}^2 m_n^2}{4\pi}.
 \label{eq:simplified}
\end{align}
The elastic scattering cross section required for the capture, in the non-relativistic limit, is given by
\begin{align}
	\sigma_{\chi n} = \frac{G_\chi^2}{\pi}\frac{m_\chi^2 m_n^2}{(m_\chi + m_n)^2}.
\end{align}
Numerically we have
\begin{align}
	&G_\mathrm{BNV} \simeq 2\times 10^{-11}\,\mathrm{GeV}^{-2} 
	\times \left(\frac{\sigma_\mathrm{BNV}v/c}{10^{-50}\,\mathrm{cm}^2}\right)^{1/2},
	\\
	&G_{\chi} \simeq 3\times 10^{-9}\,\mathrm{GeV}^{-2}
	\times \left(\frac{\sigma_{\chi n}}{10^{-45}\,\mathrm{cm}^2}\right)^{1/2},
\end{align}
for $m_\chi \gg m_n$, from which one can easily translate the constraints on $v\sigma_\mathrm{BNV}$ 
to those on $G_\mathrm{BNV}$ in Fig.~\ref{fig:results} for given $G_{\chi}$ and $m_\chi$.
As we discussed, we introduced the mass mixing $\Delta m_\chi$ to convert $\chi_2$'s back to $\chi_1$'s to catalyze baryon destruction inside NSs.
For this conversion to be efficient, we require the mass mixing time scale to be shorter than the BNV time scale inside NSs:
\begin{align}
	\Delta m_\chi \gtrsim 0.3\, \mathrm{fm}^{-3} \times \sigma_\mathrm{BNV} v/c,
\end{align}
or
\begin{align}
	\frac{\Delta m_\chi}{m_\chi} \gtrsim 6\times 10^{-28} \left(\frac{100\,\mathrm{GeV}}{m_\chi}\right)
	\left(\frac{\sigma_\mathrm{BNV}v/c}{10^{-50}\,\mathrm{cm^2}}\right).
	\label{eq:dmchi_size}
\end{align}
As we will see in the next sub-section, this condition is easily satisfied while evading terrestrial constraints on the BNV interactions.

Comparing the typical sensitivity one can derive from NS heating (Fig.\,\ref{fig:results}) with the predictions of Eq.~\eqref{eq:simplified}, one may conclude that BNV coupling constants as low as $G_\mathrm{BNV}\propto 10^{-10}\times G_F$ can be probed. While this looks impressive, we note that this level of sensitivity does not automatically mean that very high energy scales ({\em e.g.} five orders of magnitude above the weak scale) are being probed. This is because the BNV operator involving the proton field, Eq. (\ref{eq:LBNV}), is necessarily composite, involving three quark fields. In fact, the suppression from compositeness is quite significant, and a likely UV completion of Eq.(\ref{eq:LBNV}) would have to involve a combination of heavy colored particles as well as lighter neutral sub-electroweak scale fields~\cite{Fornal:2018eol}.

%%%%%%%%%%%%%%%%%%%%%%%%%%%%%%%%%%%%%%%%%%%%%%%%%%
\subsection{Di-nucleon decay \texorpdfstring{$2N \to N + e^+/\bar{\nu}$}{2N to N e+ nu}}
\label{subsec:di-nucleon}
%%%%%%%%%%%%%%%%%%%%%%%%%%%%%%%%%%%%%%%%%%%%%%%%%%
In our model, an important constraint comes from the di-nucleon decay process $2N \to N + e^+/\bar{\nu}$. The amplitude is suppressed by one mass mixing parameter $\Delta m$.
To compute this process, we go to the mass basis. The mass matrix is 
\begin{align}
	\mathcal{L}_\mathrm{mass} = -\begin{pmatrix} \bar{\chi}_1 & \bar{\chi}_2 \end{pmatrix}
	\begin{pmatrix} m_\chi & \Delta m_\chi/2 \\ \Delta m_\chi/2 & m_\chi \end{pmatrix}
	\begin{pmatrix} \chi_1 \\ \chi_2 \end{pmatrix}.
\end{align}
Redefining the fields as
\begin{align}
	\begin{pmatrix} \chi_1 \\ \chi_2 \end{pmatrix} = \frac{1}{\sqrt{2}}\begin{pmatrix} 1 & 1 \\ -1 & 1 \end{pmatrix}
	\begin{pmatrix} \chi_- \\ \chi_+ \end{pmatrix},
\end{align}
diagonalizes the mass matrix:
\begin{align}
	\mathcal{L}_\mathrm{mass} = -m_- \bar{\chi}_- \chi_- - m_+ \bar{\chi}_+ \chi_+,
\end{align}
where $m_\pm = m_\chi\pm\Delta m_\chi/2$.
In this mass basis, our toy model's Lagrangian in Eq.~\eqref{eq:Lag} becomes\footnote{
One may instead start from this Lagrangian. In this basis, there is no oscillation between $\chi_\pm$, but it does not affect our story since both $\chi_\pm$ can destroy baryons, as is clear from the second line in Eq.~\eqref{eq:BNV_mass_basis}.
}
\begin{align}
	\mathcal{L}
	&= \mathcal{L}_\mathrm{mass}
	+ G_\chi \left(\bar{\chi}_- \gamma_\mu \chi_- + \bar{\chi}_+ \gamma_\mu \chi_+\right)
	\left(\bar{p} \gamma^\mu p + \bar{n}\gamma^\mu n\right)
	\nonumber \\
	&
	+ \frac{G_\mathrm{BNV}}{2} 
	\sum_{i,j=\pm}\lambda_{ij} \bar{\chi}_i \gamma_\mu \chi_j
	\left(\overline{e^+} \gamma^\mu p + \overline{\bar{\nu}}\gamma^\mu n \right)
	+ (\mathrm{h.c.}),
	\label{eq:BNV_mass_basis}
\end{align}
where $\lambda_{--} = \lambda_{-+} = -1$ and $\lambda_{+-} = \lambda_{++} = 1$. In this basis, both the $\chi_+$ and $\chi_-$ eigenstates violate the baryon number. This does not mean, however, that $\Delta m = m_+-m_-$ can be taken to zero, while still preserving repeated BNV scattering. Due to the appearance of a coherent $|\chi_+\rangle-|\chi_-\rangle$ state in the final state of the BNV scattering, the sequence terminates, and one needs a $\Delta m$-induced decoherence, preferably satisfying condition (\ref{eq:dmchi_size}), for a repetition of the BNV scattering. In other words, in this basis, the loss of coherence  enhances the efficiency of the BNV process, unlimited by the captured DM number.

Addressing $\Delta B = 2$ processes, we notice that the cancellation happens between the $i = j$ and $i\neq j$ terms, resulting in 
$\Delta m_\chi^2$ suppression, while for $\Delta B = 1$ processes, 
the $i = j$ terms cancel with each other, resulting in $\Delta m_\chi$ suppression.

Since we take $\Delta m_\chi$ to be small, we focus on the $\Delta B = 1$ process $2N \to N + e^+/\bar{\nu}$.
The relevant amplitude is diagrammatically given by
\begin{align}
	i\mathcal{M} &=
	\begin{tikzpicture}[baseline=(c)]
	\begin{feynman}[inline = (base.c)]
		\vertex [label=\({\scriptstyle n,\,p_1}\)](n1);
		\vertex [below = 0.15 of n1] (n1d);
		\vertex [right = of n1d] (v1);
		\vertex [right = of v1] (nu1d);
		\vertex [above = 0.15 of nu1d,label=\({\scriptstyle n,\,p_3}\)] (nu1);
		\vertex [below = of v1] (v2);
		\vertex [below = 0.5 of v1] (c);
		\vertex [left = of v2] (n2d);
		\vertex [below = 0.15 of n2d,label=270:\({\scriptstyle n,\,p_2}\)] (n2);
		\vertex [right = of v2] (nu2d);
		\vertex [below = 0.15 of nu2d,label=270:\({\scriptstyle \bar{\nu},\,p_4}\)] (nu2);
		\diagram*{
		(n1) -- [fermion] (v1) -- [fermion] (nu1),
		(v1) -- [fermion, half left, edge label=\({\scriptstyle \chi_-}\)] (v2) -- [fermion, half left, edge label=\({\scriptstyle \chi_-}\)] (v1),
		(n2) -- [fermion] (v2) -- [fermion] (nu2),
		};
	\end{feynman}
	\end{tikzpicture}
	+
	\begin{tikzpicture}[baseline=(c)]
	\begin{feynman}[inline = (base.c)]
		\vertex [label=\({\scriptstyle n,\,p_1}\)](n1);
		\vertex [below = 0.15 of n1] (n1d);
		\vertex [right = of n1d] (v1);
		\vertex [right = of v1] (nu1d);
		\vertex [above = 0.15 of nu1d,label=\({\scriptstyle n,\,p_3}\)] (nu1);
		\vertex [below = of v1] (v2);
		\vertex [below = 0.5 of v1] (c);
		\vertex [left = of v2] (n2d);
		\vertex [below = 0.15 of n2d,label=270:\({\scriptstyle n,\,p_2}\)] (n2);
		\vertex [right = of v2] (nu2d);
		\vertex [below = 0.15 of nu2d,label=270:\({\scriptstyle \bar{\nu},\,p_4}\)] (nu2);
		\diagram*{
		(n1) -- [fermion] (v1) -- [fermion] (nu1),
		(v1) -- [fermion, half left, edge label=\({\scriptstyle \chi_+}\)] (v2) -- [fermion, half left, edge label=\({\scriptstyle \chi_+}\)] (v1),
		(n2) -- [fermion] (v2) -- [fermion] (nu2),
		};
	\end{feynman}
	\end{tikzpicture},
\end{align}
where we focus on $nn \to n \bar{\nu}$.
In the small mass mixing limit and $m_n^2 \ll m_\chi^2$, we obtain
\begin{align}
\label{eq:3n1l}
	i\mathcal{M} &\simeq %\frac{im_n^2}{24\pi^2}\frac{\Delta m_\chi}{m_\chi} G_\mathrm{BNV} G_\chi
	%\nonumber \\
	{\cal C}_{NNNl}\times
	\left[\bar{u}_n(p_3) \gamma_\alpha u_n(p_1)\right]
	\left[\bar{u}_{\bar{\nu}}(p_4) \gamma^\alpha u_n(p_2)\right],\\&
    {\cal C}_{NNNl}=\frac{im_n^2}{24\pi^2}\frac{\Delta m_\chi}{m_\chi} G_\mathrm{BNV} G_\chi.
\end{align}
where we note that there is a cancellation between the first and second diagrams.
Since we deal with a non-renormalizable theory, on top of the loop contribution computed above, 
we have the freedom to include higher-dimensional operators by hand that induce $nn \to n \bar{\nu}$.
In this sense, the numerical coefficient in the above should be understood as only indicative.
From this amplitude, we obtain the cross section
\begin{align}
	v\sigma(nn \to n\bar{\nu}) &= \frac{3^2}{2^{15}\pi^5}\left(\frac{\Delta m_\chi}{m_\chi}\right)^2 
	G_\mathrm{BNV}^2 G_\chi^2m_n^6.
\end{align}
Converting the cross section to the nucleon decay rate inside a nucleus in principle
requires the knowledge of the nucleon wave-function.
Instead, we may perform a simple estimate of the rate by multiplying the cross section by the nuclear density~\cite{Goity:1994dq}
\begin{align}
\label{eq:nnnnu}
	\Gamma(nn\to n\bar{\nu}) \sim 0.12 \,\mathrm{fm}^{-3}\times v\sigma (nn \to n\bar{\nu}),
\end{align}
to obtain
\begin{align}
	\tau(nn\to n\bar{\nu})
	&\sim 10^{16}\,\mathrm{yrs}
	\times \left(\frac{m_\chi}{\Delta m_\chi}\right)^2
	\nonumber \\
	&\times \left(\frac{10^{-50}\,\mathrm{cm}^2}{\sigma_\mathrm{BNV}v/c}\right)
	\left(\frac{10^{-45}\,\mathrm{cm}^2}{\sigma_{\chi n}}\right).
 \label{eq:taunn}
\end{align}
Without the two species $\chi_1$ and $\chi_2$ (or $\chi_\pm$), the rate contains a UV log divergence instead of the suppressing factors of $\Delta m_\chi/m_\chi$
and is severely constrained.

In this di-nucleon decay process, a neutron is converted to a neutrino and thus it leaves a hole in the nuclear shell.
Refilling the shell in nuclei such as $^{12}$C or $^{16}$O will result in detectable signals for leading neutrino observatories. 
Moreover, the outgoing neutron gets $\sim m_n/4$ kinetic energy and is ejected
from the nucleus. This may leave an additional signal by, e.g., the Gd capture inside SK, or via $p+n\to {\rm D} +\gamma$ reaction in various hydrocarbon-based neutrino detectors.
To the best of our knowledge, there is no experimental search looking for this specific decay mode.

Thus, we just use the invisible decay searches in Refs~\cite{Kamiokande:1993ivj,SNO:2003lol,Borexino:2003igu,KamLAND:2005pen,SNO:2018ydj}
assuming that the ejected neutron does not leave visible signals.
By requiring that $\tau(nn\to n\bar{\nu}) > 1.4\times 10^{30}\,\mathrm{yrs}$~\cite{KamLAND:2005pen}, 
we obtain an upper bound on $\Delta m_\chi/m_\chi$:
\begin{align}
\label{eq:constraint}
	\frac{\Delta m_\chi}{m_\chi} \lesssim 10^{-7}
	\left(\frac{10^{-50}\,\mathrm{cm^2}}{\sigma_\mathrm{BNV}v/c}\right)^{1/2}
	\left(\frac{10^{-45}\,\mathrm{cm^2}}{\sigma_{\chi n}}\right)^{1/2}.
\end{align}

Comparing this to Eq.~\eqref{eq:dmchi_size}, the terrestrial BNV processes can easily be avoided while having sufficiently fast $\chi_2 \to \chi_1$ conversion, or equivalently fast decoherence of the final state $\chi_\pm$. While strong mass degeneracy $\Delta m_\chi/m_\chi \ll 1$ may look like an additional fine-tuning, the universality of interaction ${\cal L}_2$ will not contribute to the mass splitting, and small $\Delta m_\chi$ will remain technically natural in the limit of small BNV processes. 

Additionally, this constraint excludes $\Delta m_\chi/m_\chi \sim \mathcal{O}(1)$, necessitating the two components, $\chi_1$ and $\chi_2$, to suppress the BNV.
On top of $nn \to n \bar{\nu}$, we can also have visible modes such as $pp \to p e^+$ and $p n \to e^+ n$
depending on the DM couplings to the protons,
but these modes do not affect our conclusion that we can avoid the terrestrial BNV constraints
by the suppression from the mass mixing. 

We would like to comment that the $nn\to n\nu$ amplitude will also lead to $n\to \pi^0\nu$ decays with {\em additional} loops. While more loops may lead to a smaller answer, the nucleon loops could effectively replace the nuclear density in Eq.~(\ref{eq:nnnnu}) by a UV scale associated with the nucleon loop, which is likely to coincide with the hadronic scale of $\sim1\,{\rm GeV}$. This may lead to event rates for the nucleon decay competitive with Eq.(\ref{eq:taunn}). A more precise comparison of nucleon decay and di-nucleon annihilation can be achieved in UV-complete models, where the underlying quark loops can be properly evaluated. For the interplay of UV completion and loop-induced BNV processes, see, {\em e.g.}, the recent work \cite{Fox:2024kda}. 

While precise answers are challenging to obtain in our incomplete framework, one can nevertheless estimate the transmutation of Eq.(\ref{eq:3n1l}) amplitude into a single nucleon decay 
$N\to l\pi $, of which the most important ones are, of course, $p \to \pi^0e^+$ and $n\to \pi^- e^+$. Integrating out a pair of nucleons in the amplitude [Eq.(\ref{eq:3n1l})] results in an operator of a type $\bar{\nu} \gamma^\alpha n \partial_\alpha \pi $.
A simple estimate carried out with the help of chiral perturbation theory suggests that a corresponding amplitude scales as 
\begin{align}
\label{eq:NDAestimate}
   {\cal C}_{N\pi l }\sim {\cal C}_{NNNl} \times \frac{\Lambda_{\rm hadr}^2}{16\pi^2 f_\pi} \sim {\cal C}_{NNNl} \times O({\rm 50\,MeV}) ,
\end{align}
where $f_\pi$ is the pion decay constant, and $\Lambda_{\rm hadr}\sim m_n$ is the maximum scale of validity of such treatment. Notice that the symmetries of the problem ensure that ${\cal C}_{N\pi l }$ is suppressed by the same parameters as ${\cal C}_{NNNl}$, notably $\Delta m$ of the $\chi$ sector. Since the resulting scale in the above estimate (\ref{eq:NDAestimate}), $\sim{\rm 50\,MeV} $ is very similar to a typical nuclear scale, the rate for the single nucleon decay is expected to be on the order of $\tau^{-1}(nn\to n\bar\nu)$, Eq. (\ref{eq:taunn}). We also note that depending on the exact model of BNV interaction, the actual amplitude may be suppressed, such as {\em e.g.} $(\bar N l) \Box \pi $, resulting in $O(100)$ suppression of a single nucleon decay. (On the other hand, experimental sensitivity to $p\to \pi^0 e^+$ will be superior to any other nucleon decay/annihilation mode). Exact analysis of the interplay between nucleon semi-annihilation, $NN\to N l$, and nucleon decay, $N\to \pi l$, may be of interest for future studies. At this point, we conclude that the single nucleon decays cannot significantly strenghten the constraint (\ref{eq:constraint}), and therefore they do not challenge the DM-induced BNV interactions, leaving the neutron star physics to be the most significant constraint. 

\section{Discussion \& Conclusions}
\label{sec:disc}
We have considered the possibility that the baryon number is broken by the interactions of DM with the SM. Since there are a large number of possibilities for DM, we have concentrated on a WIMP-like DM with mass not too far from the electroweak scale. Moreover, this BNV can occur in a variety of ways. In this paper, we have considered a ``minimal" possibility where the BNV interactions still conserve $B-L$, and therefore, have the same symmetry as the SM sphaleron processes. It is also clear, albeit not exploited in our paper, that the dark sector and BNV can be used to drive baryogenesis. One simple idea is to transfer the fermion-antifermion asymmetry from the dark sector to baryons \cite{Bringmann:2018sbs,DAgnolo:2015nbz}. But perhaps an even more appealing idea is to dynamically generate the baryon asymmetry using the DM BNV interactions after SM sphalerons stop. We plan to return to this topic in future work.

The main consequences of SM-DM BNV interactions can be analyzed in quite general terms, appealing only to the elastic and BNV cross sections. Without any elastic cross sections between DM and the nuclei, one could use neutrino telescope results to constrain DM-induced nucleon destruction. In particular, SK shows that for 10 GeV DM, BNV interactions cannot induce cross sections larger than 10$^{-49}$ cm$^2\times (c/v)$. 

If DM has even minuscule elastic scattering cross sections with nuclei, BNV interactions are very tightly constrained by NSs. Old NSs are interesting objects, where most particles are locked deep inside the Fermi sea, unable to absorb or emit heat. At that late stage of NS existence, the process of cooling occurs through the thermal radiation from the NS surface, and any additional process that releases heat in the NS volume would lead to larger surface temperatures. We apply this logic to the BNV processes on captured DM, finding extremely tight constraints on the nucleon disintegration cross section. The strongest constraints are found in the situation when the elastic scattering of DM leads to its capture by a NS, and for DM annihilation being ``switched off", perhaps by particle-antiparticle asymmetry in the dark sector. The tiniest levels of BNV cross sections can be probed that way down to values of $\sim$10$^{-59}$ cm$^2\times (c/v)$. Notably, the disintegration of baryons is allowed to continue in perpetuity, and the amount of energy release per one captured DM particle can exceed its rest mass by many orders of magnitude. This explains why old NSs can serve as a very powerful probe of the SM-DM BNV interactions, while not currently being sensitive to the process of DM capture and annihilation, as the maximum energy per annihilation process in the latter case is $2m_\chi$.

It is also clear that diagrams with virtual exchange by DM sector particles can lead to {\em bona fide} nucleon decay. To explore one such possibility, we formulated an effective DM model of two nearly degenerate fermions $\chi_1$ and $\chi_2$. The BNV parameter in this model can be traced back to the mass mixing of $\chi$'s. The amplitude for induced $nn\to n\bar\nu$ processes is proportional to $\Delta m$. Consequently, the constraints imposed by nucleon decay leave a lot of room for the BNV-induced heating of NSs at a small end of $\Delta m$. We conclude by noting that a dedicated search of nucleon semi-annihilation $nn\to n\bar\nu$ and $np\to ne^+$ is interesting to perform, and the results may improve the quality of the SK bounds inferred here. At the same time, future theoretical work is required to illustrate better the interplay between the nucleon semi-annihilation and the single nucleon decay, as well as their connection to the UV-complete models of BNV.

%%%%%%%%%%%%%%%%%%%%%%%%%%%%%%%%%%%%%%%%%%%%%%%%%%
\begin{acknowledgments}
We would like to thank Drs. J. Bramante, P. Fox, M. Hostert, T. Menzo, K. Olive and J. Zupan for useful discussions. The Feynman diagrams in this paper are drawn by \texttt{TikZ-Feynman}~\cite{Ellis:2016jkw}. 
Y.E. and M.P. are supported in part by U.S. Department of Energy Grant No. DE-SC0011842. A.R. acknowledges support from the National Science Foundation (Grant No. PHY-2020275) and  the Heising-Simons Foundation (Grant No. 2017-228).
\end{acknowledgments}

%%%%%%%%%%%%%%%%%%%%%%%%%%%%%%%%%%%%%%%%%%%%%%%%%%
\bibliographystyle{JHEP}
\bibliography{ref.bib}

\providecommand{\href}[2]{#2}\begingroup\raggedright\begin{thebibliography}{10}

\bibitem{Sakharov:1967dj}
A.~D. Sakharov, \emph{{Violation of CP Invariance, C asymmetry, and baryon
  asymmetry of the universe}},
  \href{https://doi.org/10.1070/PU1991v034n05ABEH002497}{\emph{Pisma Zh. Eksp.
  Teor. Fiz.} {\bfseries 5} (1967) 32}.

\bibitem{Kuzmin:1985mm}
V.~A. Kuzmin, V.~A. Rubakov and M.~E. Shaposhnikov, \emph{{On the Anomalous
  Electroweak Baryon Number Nonconservation in the Early Universe}},
  \href{https://doi.org/10.1016/0370-2693(85)91028-7}{\emph{Phys. Lett. B}
  {\bfseries 155} (1985) 36}.

\bibitem{Koren:2022bam}
S.~Koren, \emph{{A Note on Proton Stability in the Standard Model}},
  \href{https://doi.org/10.3390/universe8060308}{\emph{Universe} {\bfseries 8}
  (2022) 308} [\href{https://arxiv.org/abs/2204.01741}{{\ttfamily
  2204.01741}}].

\bibitem{Kaplan:2009ag}
D.~E. Kaplan, M.~A. Luty and K.~M. Zurek, \emph{{Asymmetric Dark Matter}},
  \href{https://doi.org/10.1103/PhysRevD.79.115016}{\emph{Phys. Rev. D}
  {\bfseries 79} (2009) 115016}
  [\href{https://arxiv.org/abs/0901.4117}{{\ttfamily 0901.4117}}].

\bibitem{Shelton:2010ta}
J.~Shelton and K.~M. Zurek, \emph{{Darkogenesis: A baryon asymmetry from the
  dark matter sector}},
  \href{https://doi.org/10.1103/PhysRevD.82.123512}{\emph{Phys. Rev. D}
  {\bfseries 82} (2010) 123512}
  [\href{https://arxiv.org/abs/1008.1997}{{\ttfamily 1008.1997}}].

\bibitem{Davoudiasl:2010am}
H.~Davoudiasl, D.~E. Morrissey, K.~Sigurdson and S.~Tulin, \emph{{Hylogenesis:
  A Unified Origin for Baryonic Visible Matter and Antibaryonic Dark Matter}},
  \href{https://doi.org/10.1103/PhysRevLett.105.211304}{\emph{Phys. Rev. Lett.}
  {\bfseries 105} (2010) 211304}
  [\href{https://arxiv.org/abs/1008.2399}{{\ttfamily 1008.2399}}].

\bibitem{Cui:2011qe}
Y.~Cui, L.~Randall and B.~Shuve, \emph{{Emergent Dark Matter, Baryon, and
  Lepton Numbers}}, \href{https://doi.org/10.1007/JHEP08(2011)073}{\emph{JHEP}
  {\bfseries 08} (2011) 073} [\href{https://arxiv.org/abs/1106.4834}{{\ttfamily
  1106.4834}}].

\bibitem{Bringmann:2018sbs}
T.~Bringmann, J.~M. Cline and J.~M. Cornell, \emph{{Baryogenesis from
  neutron-dark matter oscillations}},
  \href{https://doi.org/10.1103/PhysRevD.99.035024}{\emph{Phys. Rev. D}
  {\bfseries 99} (2019) 035024}
  [\href{https://arxiv.org/abs/1810.08215}{{\ttfamily 1810.08215}}].

\bibitem{Elor:2020tkc}
G.~Elor and R.~McGehee, \emph{{Making the Universe at 20 MeV}},
  \href{https://doi.org/10.1103/PhysRevD.103.035005}{\emph{Phys. Rev. D}
  {\bfseries 103} (2021) 035005}
  [\href{https://arxiv.org/abs/2011.06115}{{\ttfamily 2011.06115}}].

\bibitem{Fornal:2018eol}
B.~Fornal and B.~Grinstein, \emph{{Dark Matter Interpretation of the Neutron
  Decay Anomaly}},
  \href{https://doi.org/10.1103/PhysRevLett.120.191801}{\emph{Phys. Rev. Lett.}
  {\bfseries 120} (2018) 191801}
  [\href{https://arxiv.org/abs/1801.01124}{{\ttfamily 1801.01124}}].

\bibitem{Goldman:2019dbq}
I.~Goldman, R.~N. Mohapatra and S.~Nussinov, \emph{{Bounds on neutron-mirror
  neutron mixing from pulsar timing}},
  \href{https://doi.org/10.1103/PhysRevD.100.123021}{\emph{Phys. Rev. D}
  {\bfseries 100} (2019) 123021}
  [\href{https://arxiv.org/abs/1901.07077}{{\ttfamily 1901.07077}}].

\bibitem{Johns:2020mmo}
L.~Johns and S.~Koren, \emph{{Hydrogen Mixing as a Novel Mechanism for Colder
  Baryons in 21 cm Cosmology}},
  \href{https://arxiv.org/abs/2012.06584}{{\ttfamily 2012.06584}}.

\bibitem{Johns:2020rtp}
L.~Johns and S.~Koren, \emph{{The Hydrogen Mixing Portal, Its Origins, and Its
  Cosmological Effects}},  \href{https://arxiv.org/abs/2012.06591}{{\ttfamily
  2012.06591}}.

\bibitem{Berezhiani:2020zck}
Z.~Berezhiani, R.~Biondi, M.~Mannarelli and F.~Tonelli, \emph{{Neutron-mirror
  neutron mixing and neutron stars}},
  \href{https://doi.org/10.1140/epjc/s10052-021-09806-1}{\emph{Eur. Phys. J. C}
  {\bfseries 81} (2021) 1036}
  [\href{https://arxiv.org/abs/2012.15233}{{\ttfamily 2012.15233}}].

\bibitem{McKeen:2021jbh}
D.~McKeen, M.~Pospelov and N.~Raj, \emph{{Neutron Star Internal Heating
  Constraints on Mirror Matter}},
  \href{https://doi.org/10.1103/PhysRevLett.127.061805}{\emph{Phys. Rev. Lett.}
  {\bfseries 127} (2021) 061805}
  [\href{https://arxiv.org/abs/2105.09951}{{\ttfamily 2105.09951}}].

\bibitem{Koren:2022axd}
S.~Koren, \emph{{Cosmological Lithium Solution from Discrete Gauged B-L}},
  \href{https://doi.org/10.1103/PhysRevLett.131.091003}{\emph{Phys. Rev. Lett.}
  {\bfseries 131} (2023) 091003}
  [\href{https://arxiv.org/abs/2204.01750}{{\ttfamily 2204.01750}}].

\bibitem{Davoudiasl:2023peu}
H.~Davoudiasl, \emph{{Stellar signals of a baryon-number-violating long-range
  force}}, \href{https://doi.org/10.1103/PhysRevD.108.015023}{\emph{Phys. Rev.
  D} {\bfseries 108} (2023) 015023}
  [\href{https://arxiv.org/abs/2304.06071}{{\ttfamily 2304.06071}}].

\bibitem{Kuzmin:1983by}
V.~A. Kuzmin and V.~A. Rubakov, \emph{{On the Fate of Superheavy Magnetic
  Monopoles in a Neutron Star}},
  \href{https://doi.org/10.1016/0370-2693(83)91305-9}{\emph{Phys. Lett. B}
  {\bfseries 125} (1983) 372}.

\bibitem{Arafune:1983sk}
J.~Arafune and M.~Fukugita, \emph{{Limit on the Solar Monopole Abundance}},
  \href{https://doi.org/10.1016/0370-2693(83)90810-9}{\emph{Phys. Lett. B}
  {\bfseries 133} (1983) 380}.

\bibitem{Kolb:1984yw}
E.~W. Kolb and M.~S. Turner, \emph{{Limits from the Soft x-Ray Background on
  the Temperature of Old Neutron Stars and on the Flux of Superheavy Magnetic
  Monopoles}}, \href{https://doi.org/10.1086/162645}{\emph{Astrophys. J.}
  {\bfseries 286} (1984) 702}.

\bibitem{Callan:1982au}
C.~G. Callan, Jr., \emph{{Dyon-Fermion Dynamics}},
  \href{https://doi.org/10.1103/PhysRevD.26.2058}{\emph{Phys. Rev. D}
  {\bfseries 26} (1982) 2058}.

\bibitem{Rubakov:1982fp}
V.~A. Rubakov, \emph{{Adler-Bell-Jackiw Anomaly and Fermion Number Breaking in
  the Presence of a Magnetic Monopole}},
  \href{https://doi.org/10.1016/0550-3213(82)90034-7}{\emph{Nucl. Phys. B}
  {\bfseries 203} (1982) 311}.

\bibitem{Campbell:1990fa}
B.~A. Campbell, S.~Davidson, J.~R. Ellis and K.~A. Olive, \emph{{Cosmological
  baryon asymmetry constraints on extensions of the standard model}},
  \href{https://doi.org/10.1016/0370-2693(91)91795-W}{\emph{Phys. Lett. B}
  {\bfseries 256} (1991) 484}.

\bibitem{Campbell:1992jd}
B.~A. Campbell, S.~Davidson, J.~R. Ellis and K.~A. Olive, \emph{{On the baryon,
  lepton flavor and right-handed electron asymmetries of the universe}},
  \href{https://doi.org/10.1016/0370-2693(92)91079-O}{\emph{Phys. Lett. B}
  {\bfseries 297} (1992) 118}
  [\href{https://arxiv.org/abs/hep-ph/9302221}{{\ttfamily hep-ph/9302221}}].

\bibitem{Dreiner:1992vm}
H.~K. Dreiner and G.~G. Ross, \emph{{Sphaleron erasure of primordial
  baryogenesis}},
  \href{https://doi.org/10.1016/0550-3213(93)90579-E}{\emph{Nucl. Phys. B}
  {\bfseries 410} (1993) 188}
  [\href{https://arxiv.org/abs/hep-ph/9207221}{{\ttfamily hep-ph/9207221}}].

\bibitem{Fukugita:1986hr}
M.~Fukugita and T.~Yanagida, \emph{{Baryogenesis Without Grand Unification}},
  \href{https://doi.org/10.1016/0370-2693(86)91126-3}{\emph{Phys. Lett. B}
  {\bfseries 174} (1986) 45}.

\bibitem{Super-Kamiokande:2020bov}
{\scshape Super-Kamiokande} collaboration, K.~Abe et~al.,
  \emph{{Neutron-antineutron oscillation search using a 0.37 megaton-years
  exposure of Super-Kamiokande}},
  \href{https://doi.org/10.1103/PhysRevD.103.012008}{\emph{Phys. Rev. D}
  {\bfseries 103} (2021) 012008}
  [\href{https://arxiv.org/abs/2012.02607}{{\ttfamily 2012.02607}}].

\bibitem{Huang:2013xfa}
J.~Huang and Y.~Zhao, \emph{{Dark Matter Induced Nucleon Decay: Model and
  Signatures}}, \href{https://doi.org/10.1007/JHEP02(2014)077}{\emph{JHEP}
  {\bfseries 02} (2014) 077} [\href{https://arxiv.org/abs/1312.0011}{{\ttfamily
  1312.0011}}].

\bibitem{Davoudiasl:2011fj}
H.~Davoudiasl, D.~E. Morrissey, K.~Sigurdson and S.~Tulin, \emph{{Baryon
  Destruction by Asymmetric Dark Matter}},
  \href{https://doi.org/10.1103/PhysRevD.84.096008}{\emph{Phys. Rev. D}
  {\bfseries 84} (2011) 096008}
  [\href{https://arxiv.org/abs/1106.4320}{{\ttfamily 1106.4320}}].

\bibitem{Demidov:2015bea}
S.~V. Demidov and D.~S. Gorbunov, \emph{{Nucleon-decay-like signatures of
  Hylogenesis}}, \href{https://doi.org/10.1103/PhysRevD.93.035009}{\emph{Phys.
  Rev. D} {\bfseries 93} (2016) 035009}
  [\href{https://arxiv.org/abs/1507.05170}{{\ttfamily 1507.05170}}].

\bibitem{Super-Kamiokande:2015qek}
{\scshape Super-Kamiokande} collaboration, E.~Richard et~al.,
  \emph{{Measurements of the atmospheric neutrino flux by Super-Kamiokande:
  energy spectra, geomagnetic effects, and solar modulation}},
  \href{https://doi.org/10.1103/PhysRevD.94.052001}{\emph{Phys. Rev. D}
  {\bfseries 94} (2016) 052001}
  [\href{https://arxiv.org/abs/1510.08127}{{\ttfamily 1510.08127}}].

\bibitem{1985ApJ...296..679P}
W.~H. {Press} and D.~N. {Spergel}, \emph{{Capture by the sun of a galactic
  population of weakly interacting, massive particles}},
  \href{https://doi.org/10.1086/163485}{\emph{Astrophysical Journal} {\bfseries
  296} (1985) 679}.

\bibitem{Gould:1987ir}
A.~Gould, \emph{{Resonant Enhancements in WIMP Capture by the Earth}},
  \href{https://doi.org/10.1086/165653}{\emph{Astrophys. J.} {\bfseries 321}
  (1987) 571}.

\bibitem{Goldman:1989nd}
I.~Goldman and S.~Nussinov, \emph{{Weakly Interacting Massive Particles and
  Neutron Stars}}, \href{https://doi.org/10.1103/PhysRevD.40.3221}{\emph{Phys.
  Rev. D} {\bfseries 40} (1989) 3221}.

\bibitem{Bertone:2007ae}
G.~Bertone and M.~Fairbairn, \emph{{Compact Stars as Dark Matter Probes}},
  \href{https://doi.org/10.1103/PhysRevD.77.043515}{\emph{Phys. Rev. D}
  {\bfseries 77} (2008) 043515}
  [\href{https://arxiv.org/abs/0709.1485}{{\ttfamily 0709.1485}}].

\bibitem{Guillot:2019ugf}
S.~Guillot, G.~G. Pavlov, C.~Reyes, A.~Reisenegger, L.~Rodriguez, B.~Rangelov
  et~al., \emph{{Hubble Space Telescope Nondetection of PSR
  J2144\textendash{}3933: The Coldest Known Neutron Star}},
  \href{https://doi.org/10.3847/1538-4357/ab0f38}{\emph{Astrophys. J.}
  {\bfseries 874} (2019) 175}
  [\href{https://arxiv.org/abs/1901.07998}{{\ttfamily 1901.07998}}].

\bibitem{Raj:2024kjq}
N.~Raj, P.~Shivanna and G.~N. Rachh, \emph{{Reheated Sub-40000 Kelvin Neutron
  Stars at the JWST, ELT, and TMT}},
  \href{https://arxiv.org/abs/2403.07496}{{\ttfamily 2403.07496}}.

\bibitem{Bramante:2017xlb}
J.~Bramante, A.~Delgado and A.~Martin, \emph{{Multiscatter stellar capture of
  dark matter}}, \href{https://doi.org/10.1103/PhysRevD.96.063002}{\emph{Phys.
  Rev. D} {\bfseries 96} (2017) 063002}
  [\href{https://arxiv.org/abs/1703.04043}{{\ttfamily 1703.04043}}].

\bibitem{Bhattacharya:2023stq}
S.~Bhattacharya, B.~Dasgupta, R.~Laha and A.~Ray, \emph{{Can LIGO Detect
  Nonannihilating Dark Matter?}},
  \href{https://doi.org/10.1103/PhysRevLett.131.091401}{\emph{Phys. Rev. Lett.}
  {\bfseries 131} (2023) 091401}
  [\href{https://arxiv.org/abs/2302.07898}{{\ttfamily 2302.07898}}].

\bibitem{Kouvaris:2007ay}
C.~Kouvaris, \emph{{WIMP Annihilation and Cooling of Neutron Stars}},
  \href{https://doi.org/10.1103/PhysRevD.77.023006}{\emph{Phys. Rev. D}
  {\bfseries 77} (2008) 023006}
  [\href{https://arxiv.org/abs/0708.2362}{{\ttfamily 0708.2362}}].

\bibitem{Dasgupta:2020dik}
B.~Dasgupta, A.~Gupta and A.~Ray, \emph{{Dark matter capture in celestial
  objects: light mediators, self-interactions, and complementarity with direct
  detection}}, \href{https://doi.org/10.1088/1475-7516/2020/10/023}{\emph{JCAP}
  {\bfseries 10} (2020) 023}
  [\href{https://arxiv.org/abs/2006.10773}{{\ttfamily 2006.10773}}].

\bibitem{Baryakhtar:2017dbj}
M.~Baryakhtar, J.~Bramante, S.~W. Li, T.~Linden and N.~Raj, \emph{{Dark Kinetic
  Heating of Neutron Stars and An Infrared Window On WIMPs, SIMPs, and Pure
  Higgsinos}},
  \href{https://doi.org/10.1103/PhysRevLett.119.131801}{\emph{Phys. Rev. Lett.}
  {\bfseries 119} (2017) 131801}
  [\href{https://arxiv.org/abs/1704.01577}{{\ttfamily 1704.01577}}].

\bibitem{Raj:2017wrv}
N.~Raj, P.~Tanedo and H.-B. Yu, \emph{{Neutron stars at the dark matter direct
  detection frontier}},
  \href{https://doi.org/10.1103/PhysRevD.97.043006}{\emph{Phys. Rev. D}
  {\bfseries 97} (2018) 043006}
  [\href{https://arxiv.org/abs/1707.09442}{{\ttfamily 1707.09442}}].

\bibitem{Yakovlev:2004iq}
D.~G. Yakovlev and C.~J. Pethick, \emph{{Neutron star cooling}},
  \href{https://doi.org/10.1146/annurev.astro.42.053102.134013}{\emph{Ann. Rev.
  Astron. Astrophys.} {\bfseries 42} (2004) 169}
  [\href{https://arxiv.org/abs/astro-ph/0402143}{{\ttfamily
  astro-ph/0402143}}].

\bibitem{LZ:2022lsv}
{\scshape LZ} collaboration, J.~Aalbers et~al., \emph{{First Dark Matter Search
  Results from the LUX-ZEPLIN (LZ) Experiment}},
  \href{https://doi.org/10.1103/PhysRevLett.131.041002}{\emph{Phys. Rev. Lett.}
  {\bfseries 131} (2023) 041002}
  [\href{https://arxiv.org/abs/2207.03764}{{\ttfamily 2207.03764}}].

\bibitem{Goity:1994dq}
J.~L. Goity and M.~Sher, \emph{{Bounds on $\Delta B = 1$ couplings in the
  supersymmetric standard model}},
  \href{https://doi.org/10.1016/0370-2693(94)01688-9}{\emph{Phys. Lett. B}
  {\bfseries 346} (1995) 69}
  [\href{https://arxiv.org/abs/hep-ph/9412208}{{\ttfamily hep-ph/9412208}}].

\bibitem{Kamiokande:1993ivj}
{\scshape Kamiokande} collaboration, Y.~Suzuki et~al., \emph{{Study of
  invisible nucleon decay, N ---\ensuremath{>} neutrino neutrino anti-neutrino,
  and a forbidden nuclear transition in the Kamiokande detector}},
  \href{https://doi.org/10.1016/0370-2693(93)90582-3}{\emph{Phys. Lett. B}
  {\bfseries 311} (1993) 357}.

\bibitem{SNO:2003lol}
{\scshape SNO} collaboration, S.~N. Ahmed et~al., \emph{{Constraints on nucleon
  decay via 'invisible' modes from the Sudbury Neutrino Observatory}},
  \href{https://doi.org/10.1103/PhysRevLett.92.102004}{\emph{Phys. Rev. Lett.}
  {\bfseries 92} (2004) 102004}
  [\href{https://arxiv.org/abs/hep-ex/0310030}{{\ttfamily hep-ex/0310030}}].

\bibitem{Borexino:2003igu}
{\scshape Borexino} collaboration, H.~O. Back et~al., \emph{{New limits on
  nucleon decays into invisible channels with the BOREXINO counting test
  facility}}, \href{https://doi.org/10.1016/S0370-2693(03)00636-1}{\emph{Phys.
  Lett. B} {\bfseries 563} (2003) 23}
  [\href{https://arxiv.org/abs/hep-ex/0302002}{{\ttfamily hep-ex/0302002}}].

\bibitem{KamLAND:2005pen}
{\scshape KamLAND} collaboration, T.~Araki et~al., \emph{{Search for the
  invisible decay of neutrons with KamLAND}},
  \href{https://doi.org/10.1103/PhysRevLett.96.101802}{\emph{Phys. Rev. Lett.}
  {\bfseries 96} (2006) 101802}
  [\href{https://arxiv.org/abs/hep-ex/0512059}{{\ttfamily hep-ex/0512059}}].

\bibitem{SNO:2018ydj}
{\scshape SNO+} collaboration, M.~Anderson et~al., \emph{{Search for invisible
  modes of nucleon decay in water with the SNO+ detector}},
  \href{https://doi.org/10.1103/PhysRevD.99.032008}{\emph{Phys. Rev. D}
  {\bfseries 99} (2019) 032008}
  [\href{https://arxiv.org/abs/1812.05552}{{\ttfamily 1812.05552}}].

\bibitem{Fox:2024kda}
P.~J. Fox, M.~Hostert, T.~Menzo, M.~Pospelov and J.~Zupan, \emph{{Muon-induced
  baryon number violation}},
  \href{https://arxiv.org/abs/2407.03450}{{\ttfamily 2407.03450}}.

\bibitem{DAgnolo:2015nbz}
R.~T. D'Agnolo and A.~Hook, \emph{{Selfish Dark Matter}},
  \href{https://doi.org/10.1103/PhysRevD.91.115020}{\emph{Phys. Rev. D}
  {\bfseries 91} (2015) 115020}
  [\href{https://arxiv.org/abs/1504.00361}{{\ttfamily 1504.00361}}].

\bibitem{Ellis:2016jkw}
J.~Ellis, \emph{{TikZ-Feynman: Feynman diagrams with TikZ}},
  \href{https://doi.org/10.1016/j.cpc.2016.08.019}{\emph{Comput. Phys. Commun.}
  {\bfseries 210} (2017) 103}
  [\href{https://arxiv.org/abs/1601.05437}{{\ttfamily 1601.05437}}].

\end{thebibliography}\endgroup
%%%%%%%%%%%%%%%%%%%%%%%%%%%%%%%%%%%%%%%%%%%%%%%%%%
\end{document}